\documentclass{aa}
\usepackage{amsmath,graphics,amssymb}

\usepackage{natbib}
\bibpunct{(}{)}{,}{a}{}{,}
\defcitealias{kkii}{KKII}
\defcitealias{go}{GO}
\defcitealias{cnovit}{Paper~I}
\defcitealias{ufo}{VFKR}
\newcommand\go{\gamma}

\newcommand{\mdot}{\ensuremath{\dot M}}

\newcommand{\zav}[1]{\left(#1\right)}
\newcommand{\ms}{\ensuremath{\text{M}_{\odot}}}
\newcommand{\kms}{\ensuremath{\text{km}\,\text{s}^{-1}}}
\newcommand{\msr}{\ensuremath{\ms\,\text{year}^{-1}}} 
\newcommand{\vel}{{v}}
\newcommand{\vinfty}{\ensuremath{\vel_\infty}}
\newcommand{\hzav}[1]{\left[#1\right]}
\newcommand{\rs}{\ensuremath{R_*}}
\newcommand\de{\text{d}}
\newcommand\tlustocerv{\text{thick}\atop\text{lines}}

\newcommand{\io}[1]{{#1}_{\text{i}}}
\newcommand{\vthi}{\vel_{\text{th,i}}}

\newcommand{\vl}{{\it v}}
\newcommand{\pr}[1]{\ensuremath{{#1}_{\mathrm{p}}}}
\newcommand{\prio}[1]{\ensuremath{{#1}_{\mathrm{pi}}}}
\newcommand{\kpi}{\ensuremath{\prio{k}}}
\newcommand{\xpi}{\ensuremath{\prio{x}}}

\newcommand{\Rpi}{\ensuremath{\prio{R}}}
\newcommand{\vp}{\ensuremath{\pr{\vl}}}
\newcommand{\vi}{\ensuremath{\io{\vl}}}

\newcommand{\geff}{\ensuremath{g_\mathrm{eff}}}
\newcommand{\pderiv}[2]{\ensuremath{\frac{\partial #1}{\partial #2}}}

\allowdisplaybreaks

\begin{document}

\title{Weak wind effects in CNO driven winds of hot first stars}

\author{J.  Krti\v{c}ka\inst{1} \and V. Votruba\inst{2,1} \and J. Kub\'at\inst{2}}
\authorrunning{J. Krti\v{c}ka,  V. Votruba and J. Kub\'at}

\institute{\'Ustav teoretick\'e fyziky a astrofyziky P\v{r}F MU,
            CZ-611 37 Brno, Czech Republic, \email{krticka@physics.muni.cz}
           \and
           Astronomick\'y \'ustav, Akademie v\v{e}d \v{C}esk\'e
           republiky, CZ-251 65 Ond\v{r}ejov, Czech Republic}

\date{Received 11 December 2009}

\abstract{During the evolution of rotating first stars, which initially consisted
of only hydrogen and helium, CNO elements may emerge to their surface.
These stars may therefore have winds that are driven 
only
by
CNO elements.} {We
study weak wind effects (Gayley-Owocki heating and multicomponent effects) in
stellar winds of first generation stars driven purely by CNO elements.} {We
apply our NLTE multicomponent models and hydrodynamical
simulations.} {The multicomponent effects
(frictional heating and decoupling)
are important particularly for low
metallicity winds, but they influence mass loss
rate
only if they cause
decoupling for velocities lower than the escape velocity. The multicomponent
effects also modify the feedback from first stars. As a result of the decoupling of
radiatively accelerated metals from hydrogen and helium, the first low-energy cosmic
ray particles are generated. We study the interaction of these particles with
the interstellar medium concluding that these particles easily penetrate the
interstellar medium of a given minihalo. We discuss the charging of the first
stars by means of
their winds.} {Gayley-Owocki heating, frictional heating, and the decoupling of
wind components occur in the winds of evolved low-metallicity stars and the
solar metallicity main-sequence stars.}


\maketitle

\section{Introduction}

Many aspects of chemical evolution in our Universe remain unclear. 
The first elements, helium and a very small amount of, e.g.,~lithium,
were 
almost certainly synthesised during the era of primordial nucleosynthesis
\citep[e.g.,][]{coc}. The first stars in the Universe can therefore be
considered to be purely hydrogen-helium stars \citep[see][for a
review]{lesdiablerets}.

Subsequent chemical evolution is less clear, partly because it is difficult
to test the theoretical predictions observationally. 
For example,
stars with a very
low abundance of iron are observed,
which
are expected to be relics from
ancient times \citep[e.g.,][]{starka}. However, their relevance as observational
testbeds to the theory of evolution of chemical composition may be hampered by
several secondary effects \citep{torna}. 
Several
hypotheses
have been developed to explain the
chemical composition of these stellar relics
\citep[cf.,][]{sty,medaci,mee}.

Hot star winds are supposed to play
an
important role in
the
chemical evolution of our
Universe. Since they remove material from the outer stellar envelopes, they only affect
the stellar mass
during the early phases of stellar evolution and do not
contribute to the change in the chemical composition of the interstellar medium. On
the other hand, as soon as freshly synthesised elements emerge at the stellar
surface during later phases of stellar evolution \citep{mee,samhir}, the hot
star winds may contribute to the chemical evolution of the interstellar medium even
before the star possibly explodes as a supernova.

Hot star winds are studied mainly by assuming a solar mixture of elements and
information about winds of more exotic composition is scarce
\citep{vikowr,grahamz,uncno}.
An interesting mixture of heavier elements, which
is
uncommon in 
contemporary
Universe, is represented by a pure CNO composition. This
composition may
be typical 
of
later phases in the evolution of the first stars. This is connected to
the possibility that the envelopes of the first stars in later evolutionary
phases are
enriched by the products of helium burning \citep{mee,samhir}.
%
%
A chemical mixture rich in CNO elements and underabundant in iron is typical
for one group of low-metallicity stars \citep[e.g.,][]{pannorris,starka}.

The study of CNO driven winds is important not only for early stellar generations
\citep[cf.,][]{uncno}. The low density winds of present stars are also accelerated
mostly by CNO lines because the contribution of other heavier elements is relatively
small \citep[e.g.,][]{vikolamet}.

To understand the role of CNO driven winds in hot evolved first stars,
\citet[hereafter \citetalias{cnovit}]{cnovit} calculated wind models of these
stars. They concluded that CNO elements do not drive winds as efficiently as iron
peak elements because of the lower number of their strong lines. Therefore, the total
amount of mass lost 
%
by these winds 
%
does not significantly
affect stellar evolution.  For subsequent stellar generations, the wind
enrichment of primordial halos by heavier elements does not 
%
overcome the
metallicity threshold for the formation of very massive stars. On the other hand, the
enrichment could be large enough to change the behavior of primordial stars
during their formation -- a mass fraction of CNO higher than about $10^{-10}$ is
sufficient to enable hydrogen burning via CNO cycle and preclude initial helium
burning by means of the 3$\alpha$ reaction \citep{vittorio}.

Some  CNO driven winds (especially the low-metallicity ones) may be subject to
weak wind effects. For weak winds, two effects that are negligible for high-density winds
may become important, namely the Gayley-Owocki (Doppler) heating/cooling
\citep[hereafter \citetalias{go}]{go} and multicomponent effects 
\citep[hereafter \citetalias{kkii}]{kkii}.

The GO heating/cooling is caused by a frequency difference between photons
entering and escaping the Sobolev resonance zone. Multicomponent effects are
connected to the momentum transfer between heavier elements (accelerated by
line absorption) and bulk wind material, i.e., hydrogen and helium. In
low-density winds, momentum transfer may become inefficient, causing frictional
heating or even decoupling of wind components
\citep{cak76,treni,kkii,op,ufo,uncno}.

To understand the role of weak wind effects in the CNO winds of massive first
stars, we calculate models of the multicomponent winds of these stars for which GO
heating is taken into account.

\section{Model assumptions}

The parameters of the model stars studied (see Table~\ref{hvezpar}) were obtained from
the grid of the evolutionary models of initially zero-metallicity stars
calculated by \citet{bezmari}. The parameters of these model stars were selected
to cover a large area of the HR diagram. Only stars that may have wind for the
mass fraction of heavier elements $Z\leq10^{-2}$ are considered here
\citepalias[see][]{cnovit}. 

For the study of a multicomponent stellar wind, we used two different codes. We
first
used a NLTE multicomponent stationary wind code
\citep[\citetalias{cnovit}]{kkii}, briefly described in Sect.~\ref{krtekmod}.
This code enables
us
to predict a wind mass-loss rate and hydrodynamical structure,
but does not allow
us
to study the evolution of the decoupled flow. Second, we used
a time-dependent multicomponent code \citep{ufo} to study the decoupling of wind
components (see Sect.~\ref{ufomod}).

\subsection{NLTE multicomponent wind models}
\label{krtekmod}

\begin{table}[t]
\caption{Radius $R_*$, mass $M$, and effective temperature $T_\text{eff}$
of studied model stars}
\label{hvezpar}
\begin{tabular}{ccccc@{}cccc}
\multicolumn{4}{c}{Evolved stars} & &
\multicolumn{4}{c}{ZAMS stars}\\
\cline{1-4}
\cline{6-9}\\[-3mm]
\cline{1-4}
\cline{6-9}
Model & $R_*$ & $M$ & $T_\text{eff}$ & & Model & $R_*$ & $M$ & $T_\text{eff}$ \\
& $[\text{R}_\odot]$ & $[\text{M}_\odot$] & [kK] & &
& $[\text{R}_\odot]$ & $[\text{M}_\odot$] & [kK] \\
\cline{1-4}
\cline{6-9}
M999-1 & 56.4 & 100 & 29.9 & & M200& 1.65 &  20 & 65.3 \\
M999-2 &125   & 100 & 20.1 & & M120& 1.42 &  12 & 49.9 \\
M999-3 &510   & 100 & 10.0 & & M070& 1.30 &   7 & 34.8 \\
M500-1 & 11.1 &  50 & 50.0 & & M040& 1.17 &   4 & 23.6 \\
\cline{6-9}
M500-2 & 33.7 &  50 & 29.9 \\
M500-3 & 72.0 &  50 & 20.6 \\
M500-4 & 303  &  50 & 10.1 \\
M200-1 &  4.1 &  20 & 50.0 \\
M200-2 & 19.9 &  20 & 24.5 \\
M100-1 & 11.4 &  10 & 20.2 \\
M100-2 & 45.6 &  10 &  9.8 \\
M050-1 &  5.1 &   5 & 20.1 \\
\cline{1-4}
\end{tabular}
\end{table}

A reader interested in a more detailed description of the latest version of our
NLTE multicomponent wind code can consult \citetalias{cnovit} for more detailed
information on NLTE equations and 
\citetalias{kkii}
for description of multicomponent
hydrodynamic equations. Here we only summarize the basic features of our models.

To calculate NLTE multicomponent wind models, we assume a spherically
symmetric stationary stellar wind. The excitation and ionization state of considered
elements is derived from the statistical equilibrium (NLTE) equations. The ionic
models are either taken from the OSTAR2002 grid of model stellar atmospheres
\citep{ostar2003,bstar2006} or prepared by ourselves. The ionic models used here are
based mainly on the Opacity Project data \citep{topp,top1,toptu,topt,topf}. For
more details of our ionic list, we refer to \citetalias{cnovit}.

The solution of the radiative transfer equation is simplified
for
%
both continua
(neglect of line transitions) and lines (using the Sobolev approximation,
\citealt{cassob}). The line radiative force is calculated in the Sobolev
approximation using NLTE level populations.
The Sobolev approximation and in particular the critical point approach for
the calculation of mass-loss rates has gained some criticism
\citep{lucyinthewind,divnetomaj}. However, we note that the comoving-frame
calculation of the radiative force agrees with the
Sobolev
approximation in the
supersonic part of smooth line-driven winds \citep{ppk,pulpren}. Moreover,
the mass-loss rates derived from hydrodynamical simulations agree with
those
derived using CAK critical point approach \citep{ocr,felpulpal}.
Atomic data for the line radiative force
calculations are taken from the VALD database \citep{vald1,vald2}, and also partly
from \citet{nist} and \citet{kur01}. 
The surface emergent flux (i.e., the lower boundary condition for the radiative
transfer in
the
wind) is taken from the H-He spherically symmetric NLTE model stellar
atmospheres of \citet[and references therein]{kub}.

The derived radiative force (including the force due to the light scattering on
free electrons) is used to solve the hydrodynamic equations. We solve the
equation of continuity, momentum, and energy equations for each component of the
flow. To calculate the radiative cooling/heating term, we use the thermal
balance of electrons method \citep{kpp} taking 
all considered bound-bound, bound-free, and free-free transitions
into account.

For our present purposes we calculated five-component wind models with wind
components corresponding to carbon, nitrogen, oxygen, and free electrons, and
a passive wind component (hydrogen
and helium). The inclusion of GO heating into our models is described in
Sect.~\ref{gokapri}.

\subsection{Time-dependent multicomponent wind models}
\label{ufomod}

To calculate time-dependent models, we restrict ourselves to a 1D spherically
symmetric, isothermal, quasineutral, two-component outflow consisting of metals
(namely oxygen, carbon, and nitrogen) that scatter stellar photons in numerous
spectral lines, and passive plasma (consisting of hydrogen and helium). We use
the simplified two-component model instead of a more accurate five-component one
because we wish to suppress numerical instabilities and keep the problem
finitely computable. The acting forces in our model are gravity, dynamical friction,
gas pressure gradients, and the line radiative force, the last of which acts only
on line-scattering ions.

The continuity equations are \citep[see Eqs.\,(1) and (2) of][]{ufo}
\begin{subequations}
\label{kont}
\begin{eqnarray} 
\pderiv{\pr\rho}{t}+
\frac{1}{r^2}\pderiv{(r^2 \pr\rho \vp)}{r}&=&0\,, \\
\pderiv{\io\rho}{t}+
\frac{1}{r^2}\pderiv{(r^2 \io\rho \vi)}{r}&=&0\,,
\label{kontm}
\end{eqnarray}
\end{subequations}
and the equations of motion are
\begin{subequations}
\label{momentum}
\begin{eqnarray}
\pderiv{\vp}{t}+\vp\pderiv{\vp}{r}+\frac{1}{\pr\rho}\pderiv{\pr{p}}{r}
	&=&\frac{\Rpi}{\pr\rho}-\geff \,, \\		
\pderiv{\vi}{t}+\vi\pderiv{\vi}{r}+\frac{1}{\io\rho}\pderiv{\io{p}}{r}
	&=& g_{\rm rad}^\text{i}-\geff-\frac{\Rpi}{\io\rho} \,.
\end{eqnarray}
\end{subequations}
In these equations, ${\io\rho}, {\vi}$, and $\io{p}$ represent the density, velocity, and
pressure of metals, respectively, and ${\pr\rho},{\vp}$, and $\pr{p}$ denote the
same quantities for passive plasma. The line radiative force is denoted by
$g_{\rm rad}^\text{i}$. The effective gravitational acceleration is $\geff =
GM(1-\Gamma_e)/r^2$, where $\Gamma_e$ is the Eddington factor, i.e., the ratio
of radiative force caused by 
electron scattering to gravitational force and $G$ is the gravitational
constant. In this two-component model, radiative force caused by Thomson
scattering acts on both components, absorbing ions and passive plasma
\citep{op}. The frictional force {\Rpi} between metals and passive plasma is
expressed as
\begin{equation}
\label{trsila}
\Rpi=\pr{n} \io{n} \kpi G(\xpi)\,,
\end{equation}
where $\pr{n}$, $\io{n}$ are the number densities of passive plasma and
absorbing ions, respectively, $\kpi$ is the frictional coefficient,
$G(\xpi)$ is the Chandrasekhar function, and $\xpi$ is the dimensionless
drift speed between components \citep[see][Eqs.\,(11) and (12)]{ufo}.

To solve the hydrodynamic equations Eqs.~\eqref{kont} and \eqref{momentum}, we
use the hydrodynamic code described in \citet{ufo}, which provides technical details of the
code and describes the numerical schemes used. Here we changed the method of
calculation of  the frictional term. Dynamical friction is the most difficult
term to compute. 
Due to big stiffness of the system of partial differential
equations, we apply a fully implicit scheme for this term \citep[see][]{ufo2}.

We adopt a relatively small Courant number $0.05$, which can stabilize the
numerical problems caused by strong decoupling instability. This instability is
caused by the dependence of the frictional force on the drift velocity between
species (for more details about the decoupling instability, see \citealt{op},
\citealt{kkiii}).
As the final Courant time step, we use the minimum of the separate time steps
calculated for both individual flow components. 

\section{%
GO
heating}

\subsection{Inclusion in the NLTE models}
\label{gokapri}

The Gayley-Owocki (GO, Doppler)
heating term (per unit of volume) is given in the
Sobolev approximation by summation of heating contributions over all lines
\citep[see Eq.~(28) therein]{go,kkii}
\begin{equation}
\label{cakgojed}
\io Q^{\text{GO}} = \frac{2\pi\vel}{rc^2}
\sum_{\text{lines}}\vthi\nu_{ij} I_\text{c} \,\go\! \zav{t_{ij},\sigma,\mu_*},
\end{equation}
where $v$ is the radial wind velocity, the thermal speed of ion with mass
$m_\text{i}$ is $\vthi=\zav{2kT/m_\text{i}}^{1/2}$, $I_\text{c}=4H_\text{c}$ is the core
intensity (frequency-dependent), $t_{ij}=\chi_{ij}cr/\zav{\nu_{ij}\vel}$,
\begin{multline}
\label{gfce}
\go (t,\sigma,\mu_*)=t\int_{-1}^1 \de\mu
\hzav{D(\mu)-\frac{\beta_\text{c}}{\beta}}\times\\*\times
\int_{-\infty}^{\infty} \de x\, x \phi(x)
\exp\zav{-\frac{t\Phi(x)}{1+\sigma\mu^2}},
\end{multline}
the frequency-integrated line opacity is
\begin{equation}
\label{opac}
\chi_{ij}=\frac{\pi e^2}{m_\text{e}c}
\zav{\frac{n_i}{g_i}-\frac{n_j}{g_j}} g_if_{ij},
\end{equation}
where $n_i$, $n_j$, $g_i$, and $g_j$ are number densities and statistical
weights of levels giving rise to the line with oscillator strength $f_{ij}$ and
frequency $\nu_{ij}$, $D(\mu)$ is unity for $\mu>\mu_*$ and zero otherwise
($\mu_*=\zav{1-R_*^2/r^2}^{1/2}$), core penetration and escape probabilities are
given by
\begin{subequations}
\label{betka}
\begin{align}
\label{betac}
\beta_\text{c}&=\frac{1}{2}\int_{\mu_*}^{1}
\frac{1-\exp\zav{-\frac{t}{1+\sigma\mu^2}}}{\frac{t}{1+\sigma\mu^2}}\de \mu\\
\intertext{and}
\label{beta}
\beta&=\frac{1}{2}\int_{-1}^{1}
\frac{1-\exp\zav{-\frac{t}{1+\sigma\mu^2}}}{\frac{t}{1+\sigma\mu^2}}\de\mu,
\end{align}
\end{subequations}
respectively,  $\phi(x)$ is the line profile (assumed to be given by a Gaussian
function),
\begin{equation}
\Phi(x)=\int_{x}^{\infty}\de x'\phi(x'),
\end{equation}
and the variable $\sigma$ was introduced by \cite{cassob} to be
\begin{equation}\label{sigma}
\sigma=\frac{r}{\vel}\frac{\de \vel}{\de r} -1.
\end{equation}
This variable determines the sign of the GO heating. For $\sigma>0$, the sign of the
function $\go(t,\sigma,\mu_*)$ is negative, causing the wind GO cooling
(typically close to the star). On the other hand, for $\sigma<0$ 
the sign is positive, which
corresponds to
heating.

To calculate the function $ \go (t,\sigma,\mu_*)$, we use the numerical
quadrature \citepalias{kkii}. First, the integral over $x$ can be efficiently
computed using a Hermite quadrature formula. Quadrature weights and knots were
computed using the subroutine {\tt  IQPACK}, which is an implementation of a
method described by \citet{iq}. A satisfactory approximation can be obtained using
50 quadrature points. For large $t$ ($\gtrsim100$), the Hermite quadrature
formula becomes inefficient and we use the simple trapezoidal rule. For angle
integration, we used the Legendre quadrature formula with 5 quadrature points. 
Quadrature weights and knots were again computed using the subroutine {\tt  IQPACK}
\citep{iq}.

To ensure the convergence of the model equations, we also included the derivatives of the GO
heating term Eq.~\eqref{cakgojed} with respect to the corresponding model
variables,
in the Newton-Raphson iteration step.

\subsection{The effect of GO heating}
\label{vlivgo}

To understand the role of GO heating in low density winds, we first neglect the
frictional heating. This enables us to compare the magnitude of GO heating with
the competing cooling processes, namely radiative and adiabatic cooling. The
cooling processes
are included consistently in our NLTE models, although in the following we
use optically thin radiative cooling after \citet{rcs},
\begin{equation}
\label{zaroch}
Q^\text{rad}\approx n_\text{H}n_\text{e}\Lambda(T)\approx
n_\text{H}^2\Lambda(T),
\end{equation}
and adiabatic cooling
\begin{equation}
\label{adoch}
{Q^{\text{ad}}}= {a^2\rho \frac{1}{r^2}\frac{\text{d}}{\text{d}r}
\zav{r^2 {\vel}}},
\end{equation}
where $\Lambda(T)$ is the cooling function. For simplicity, we assume that
$n_\text{H}\approx n_\text{e}$, $a^2\approx2kT/m_\text{H}$.

\begin{figure}
\centering
\resizebox{0.9\hsize}{!}{\includegraphics{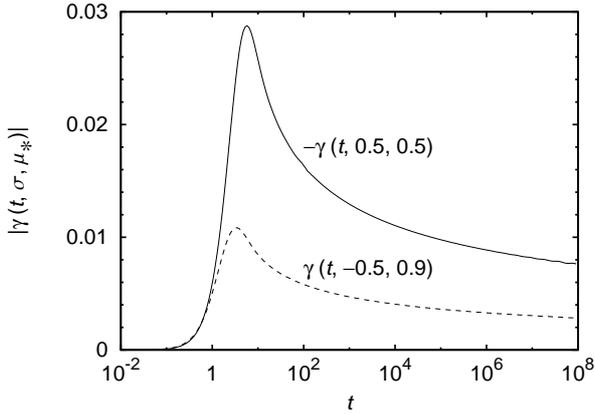}}
\caption{The dependence of $ \go (t,\sigma,\mu_*)$ on $t$ for selected
specific values of $\sigma$ and $\mu_*$}
\label{kresg}
\end{figure}

\subsubsection{Low-density winds}
\label{malhust}

The size of the GO heating/cooling term may be inferred from the plot of
the function $ \go (t,\sigma,\mu_*)$ in Fig.~\ref{kresg}. For winds of low
density, the line optical depths are small, ($t/(1+\sigma\mu_*^2)\lesssim1$),
and we can use the Taylor expansion of the exponentials in
Eqs.~\eqref{gfce}, \eqref{betka}. Assuming the line profile given by the
Gaussian function, we derive for $\sigma>0$ the approximation
\begin{equation}
\label{gten}
\go (t,\sigma,\mu_*)\approx\frac{t^2}{\sqrt{8\pi\sigma}}
\hzav{\mu_*\arctan\sqrt\sigma-\arctan\zav{\mu_*\sqrt\sigma}}
\end{equation}
and in a similar way also for $\sigma<0$. Because it follows from Eqs.~\eqref{cakgojed} --
\eqref{opac} that $t\sim\rho$, 
%
the GO heating/cooling term in Eq.~\eqref{gten} 
is proportional
to $\rho^2$.
On the other hand, the adiabatic cooling term 
in Eq.~\eqref{adoch} is linearly
proportional to density. 
Therefore,
for very low wind densities when all
lines become optically thin, adiabatic cooling begins to dominate and GO heating
is not important.

\subsubsection{High-density winds}

From Fig.~\ref{kresg}, it follows that for strong lines $t\gg1$
%
the GO
heating function $\go(t,\sigma,\mu_*)$ does not significantly depend on $t$.
The value of $\io Q^{\text{GO}}$ for optically thick lines therefore
basically depends on the number of these lines.

The importance of GO optically thick heating can be inferred by its comparison with
other effects influencing temperature, namely adiabatic cooling and radiative
heating. Approximating the derivative in Eq.~\eqref{adoch} by using the velocity
law $\vel=\vinfty(1-R_*/r)$, assuming $r\gg R_*$, and $\vthi/a\approx 1/4$, then
the ratio of GO heating and adiabatic cooling is from Eqs.~\eqref{cakgojed},
\eqref{adoch}
\begin{equation}
\label{goadtl}
\frac{\io Q^{\text{GO}}}{Q^{\text{ad}}}\approx4\pi^2
\go_\text{th}\frac{\vinfty}{a}
\frac{r^2}{\mdot c^2}\sum_{\tlustocerv}\nu_{ij} H_\text{c},
\end{equation}
where $\vinfty$ is the wind terminal velocity, $\go_\text{th}$ is the value of
$\go(t,\sigma,\mu_*)$ for $t\gg1$ (typically according to Fig.~\ref{kresg}
$\go_\text{th}\approx0.01$), and $\mdot$ is the wind mass-loss rate. In scaled
quantities, Eq.~\eqref{goadtl} reads
\begin{multline}
\label{goadtlsk}
\frac{\io Q^{\text{GO}}}{Q^{\text{ad}}}\approx0.3
\frac{\zav{\displaystyle\frac{\vinfty}{10^3\,\text{km}\,\text{s}^{-1}}}}
     {\zav{\displaystyle\frac{a}{10\,\text{km}\,\text{s}^{-1}}}}
\frac{\zav{\displaystyle\frac{r}{100\, \text{R}_\odot }}^2}
{\zav{\displaystyle\frac{\mdot}{10^{-7}\,\msr}}}\\*
\times\sum_{\tlustocerv}
\zav{\frac{\nu_{ij}}{10^{15}\,\text{s}^{-1}}}
\zav{\frac{H_\text{c}(\nu_{ij})}{10^{-3}\,\text{erg}\,\text{cm}^{-2}}}.
\end{multline}
For stars with large radii ($\rs\gtrsim10 \,\text{R}_{\sun}$) and sufficiently
high number of optically thick lines, GO heating may be comparable 
to
the
adiabatic cooling and may influence wind temperature. This was  already found
by GO, as Eq.~\eqref{goadtlsk} corresponds to Eq.~(28) of GO.

A comparison of GO heating given by Eq.~\eqref{cakgojed} with radiative cooling
given by Eq.~\eqref{zaroch} implies that
\begin{equation}
\frac{\io Q^{\text{GO}}}{Q^\text{rad}}\approx
\frac{128\pi^3r^3v^3m_\text{H}^2 \gamma_\text{th}}{c^2\Lambda(T)\mdot^2}
\sum_{\tlustocerv}\vthi\nu_{ij} H_\text{c},
\end{equation}
or, assuming $\Lambda(T)=5\times10^{-23}(T/10^5\,\text{K})$ \citep{rcs} and
approximating $v\approx v_\infty$ in scaled quantities
\begin{multline}
\label{gozartlsk}
\frac{\io Q^{\text{GO}}}{Q^\text{rad}}\approx
0.07\zav{\frac{r}{100\, \text{R}_\odot }}^3
\zav{\frac{\vinfty}{10^3\,\text{km}\,\text{s}^{-1}}}^3
\zav{\frac{T}{10^4\,\text{K}}}^{-1/2}\\*
\zav{\!\frac{\mdot}{10^{-7}\,\msr}\!}^{-2}\!\!
\sum_{\tlustocerv}\!
\zav{\!\frac{\nu_{ij}}{10^{15}\,\text{s}^{-1}}\!}\!
\zav{\!\frac{H_\text{c}(\nu_{ij})}{10^{-3}\,\text{erg}\,\text{cm}^{-2}}\!},
\end{multline}
which is similar in value to Eq.~(31) of GO. If there is a sufficiently high
number of optically thick lines, then GO heating may be comparable 
to the
radiative heating in the outer regions of hot star winds.

\subsection{Numerical results}

\begin{figure}
\centering
\resizebox{0.9\hsize}{!}{\includegraphics{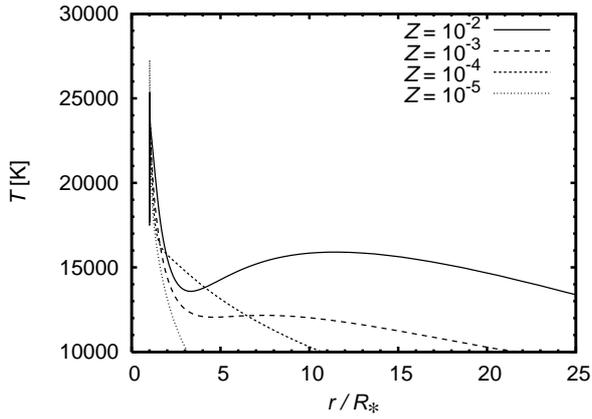}}
\caption{The effect of GO heating in the NLTE wind model \mbox{M999-1} (see
Table\,\ref{hvezpar}) obtained for different mass fractions of heavier elements.
Multicomponent effects were neglected here.}
\label{999_1t}
\end{figure}

An example of the numerical results obtained for the NLTE model M999-1 (see
Table\,\ref{hvezpar}) for different mass fractions of heavier elements $Z$ is
given in Fig.~\ref{999_1t}. The effect of the GO heating is the strongest for
the model with the highest metallicity (the highest mass-loss rate), whereas for
low metallicity models these effects are negligible.

The trends displayed in Fig.~\ref{999_1t} can be explained using the discussion
presented in Sect.~\ref{vlivgo}. The dominant cooling mechanism for stars with
a very low mass-loss rate (very low metallicity) is the adiabatic cooling (see
Sect.~\ref{malhust}). For a star with $Z=10^{-5}$, the GO heating
does not significantly influence the wind temperature. For stars with higher
mass-loss rates (higher metallicity), 
the radiative and GO processes in addition to the adiabatic cooling influence
the wind temperature. For example for
$Z=10^{-3}$ and $r=10\rs$, we can derive from Eqs.~\eqref{goadtlsk} and
\eqref{gozartlsk} using corresponding wind parameters \citepalias{cnovit} that the ratio of GO heating to both adiabatic cooling and
radiative heating is of the order of unity. The GO heating may therefore
influence wind temperature. For $Z=10^{-2}$, the mass-loss rate increases, but on
the other hand both the terminal velocity and the number of optically thick
lines increase, enhancing the importance of the GO heating.

Although the GO heating may be important for temperature balance, in the
calculations presented here
it never heats the wind to temperatures
significantly
higher than
the
stellar
effective temperature. The relative importance of this effect for temperature
balance of the wind of studied evolved stars is connected to their large radii
and in some cases also with large wind terminal velocities. For 
present-day
hot
stars with smaller radii, lower wind terminal velocities, and iron as a
significant wind driver the effect is less important, i.e., as iron lines become
optically thin in outer wind regions, their contribution to GO heating is likely
to be relatively small
(see also \citealt{vikolabis}, \citealt{pusle}).

\section{Multicomponent models}

After discussing the role of GO heating separately, we present here detailed
NLTE five-component wind models (described in Sect.~\ref{krtekmod}) with all
relevant heating/cooling effects included, i.e., radiative, adiabatic, GO, and
frictional ones. These models are supplemented by hydrodynamical simulations of
two-component flow (see Sect.~\ref{ufomod}).

\subsection{High density winds}

\begin{figure}
\centering
\resizebox{0.9\hsize}{!}{\includegraphics{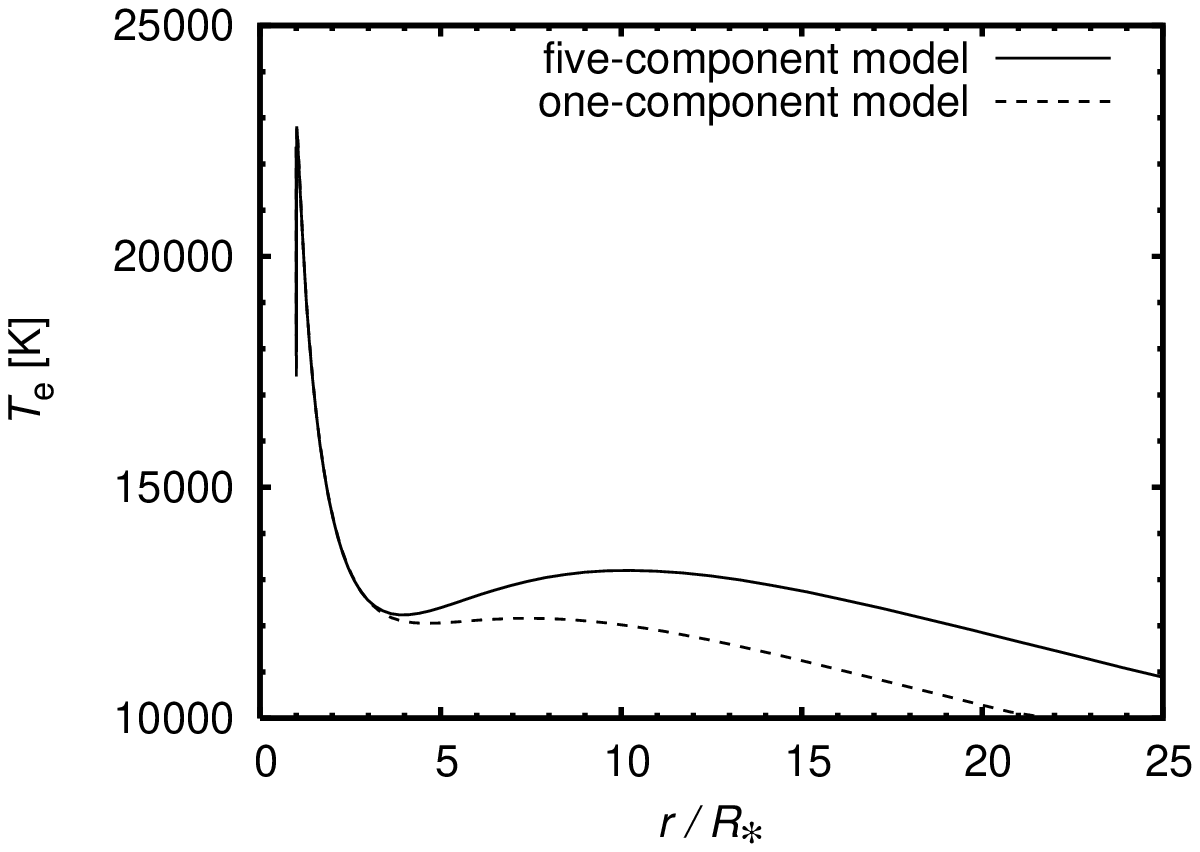}} 
\resizebox{0.9\hsize}{!}{\includegraphics{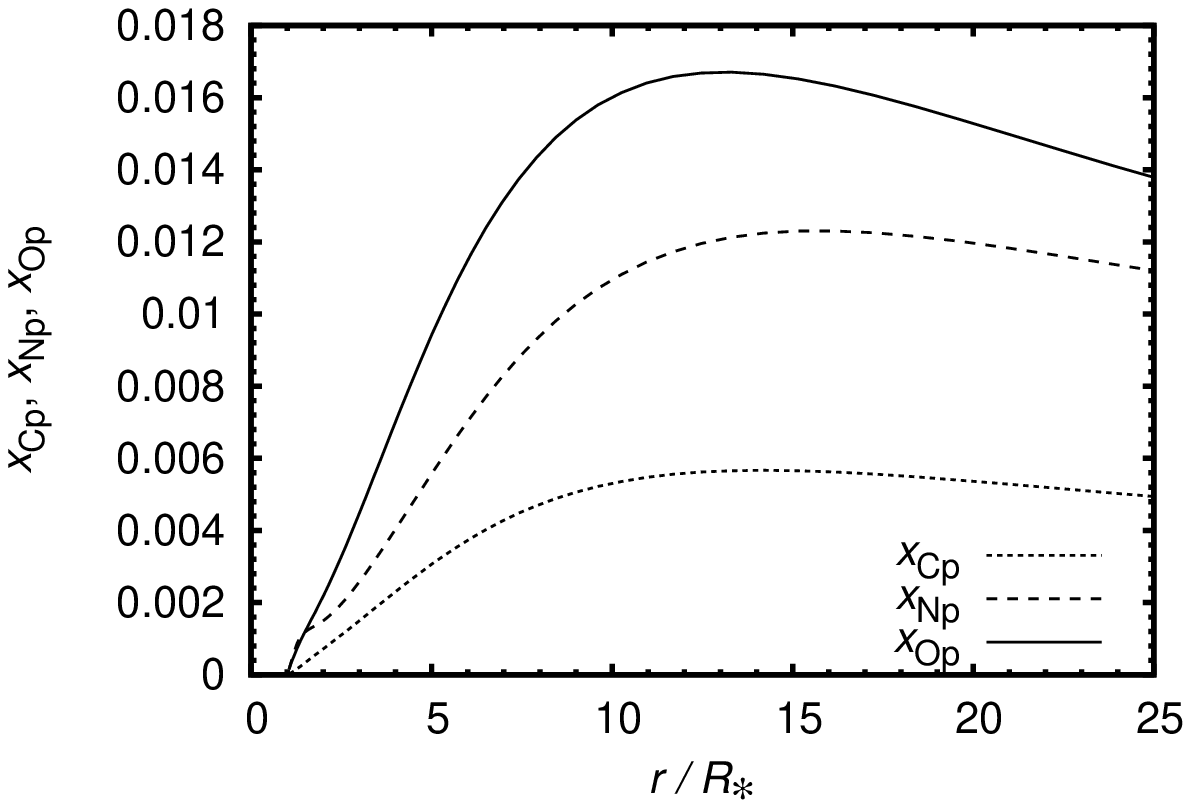}} 
\caption{Multicomponent effects in the NLTE wind model M999-1 (see
Table\,\ref{hvezpar}) for $Z=10^{-3}$. {\em Top}: Comparison of electron
temperature in the five-component and one-component wind models. {\em Bottom}:
Non-dimensional velocity difference
(Eq.~\eqref{xpi})
between passive component p
(hydrogen and helium) and carbon, nitrogen, and oxygen in the five-component
wind model.}
\label{999_1Z1e3}
\end{figure}

For stars with high density winds, the velocity differences between individual
wind components are much smaller 
than corresponding mean thermal speed.
For these stars, frictional heating is negligible and decoupling
does not occur \citep{cak76,treni,kkii,ufo}. The winds of these stars can be
adequately described by one-component models (e.g., Fig.~\ref{999_1Z1e3}).

\subsection{Winds with frictional heating}

For winds with lower densities, velocity differences between individual wind
components become comparable to the mean thermal speed, and momentum transfer
between the wind components becomes inefficient. This may cause frictional
heating of the wind \citep{treni,curdi,kkii}.

\begin{figure}
\centering
\resizebox{0.9\hsize}{!}{\includegraphics{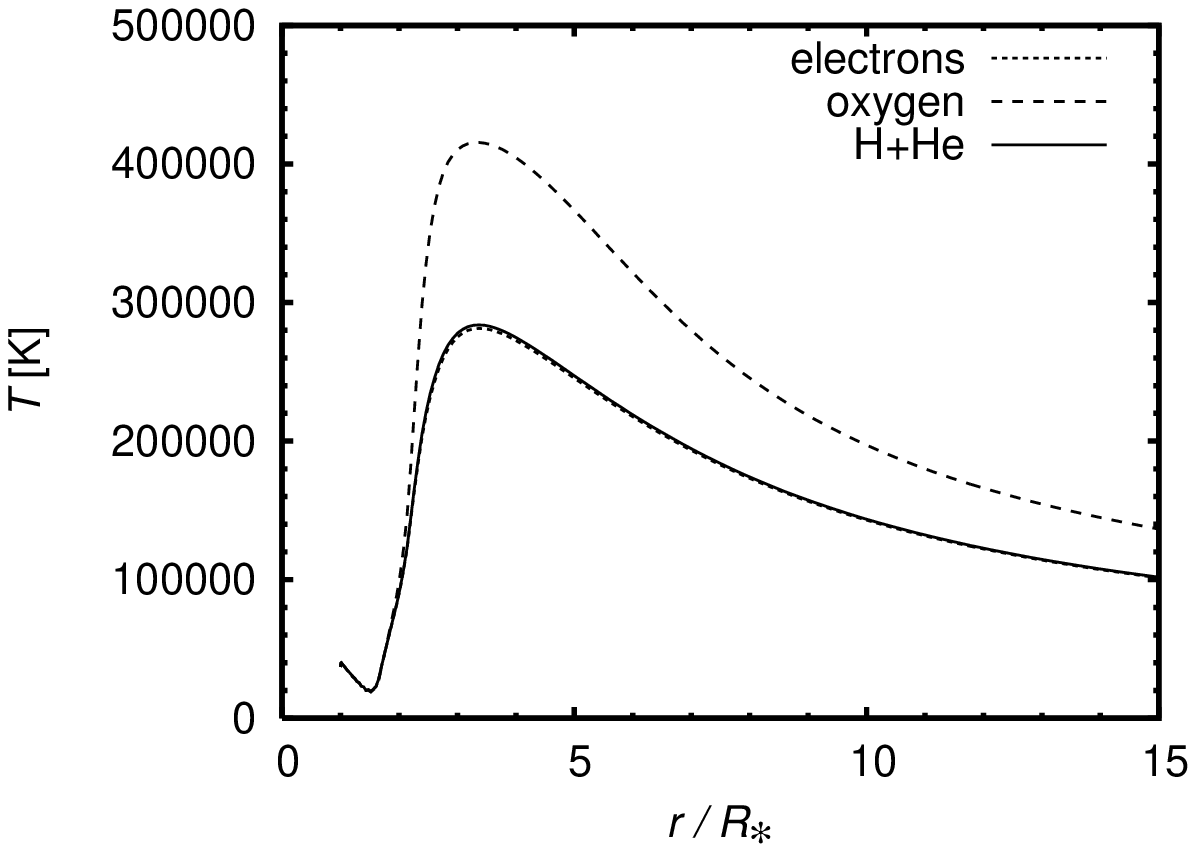}}
\resizebox{0.9\hsize}{!}{\includegraphics{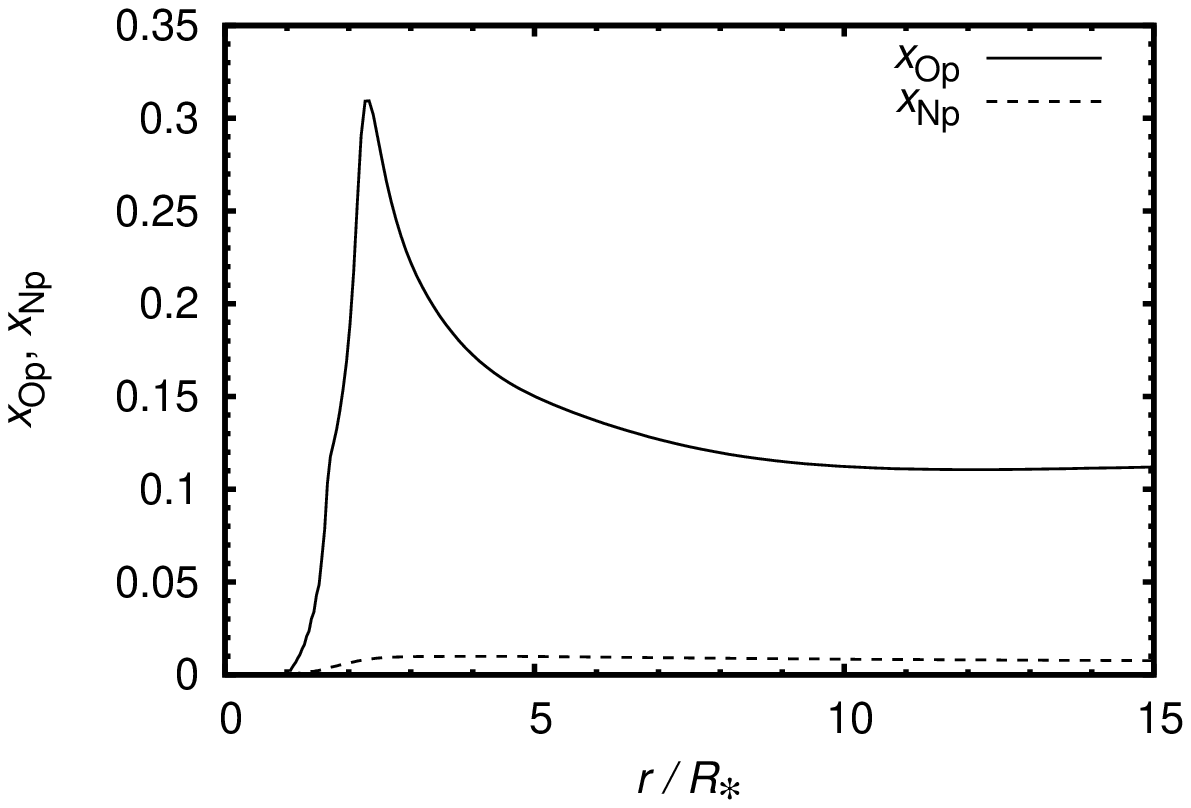}}
\caption{Frictional heating in the NLTE five-component wind model M200-1 (see
Table\,\ref{hvezpar}) for $Z=10^{-2}$. {\em Top}: Temperature of individual wind
components (oxygen, hydrogen and helium
component,
and free electrons). Temperature of
carbon and nitrogen components is nearly the same as the hydrogen and helium
one. {\em Bottom}: Nondimensional velocity difference between passive component
p (hydrogen and helium) and nitrogen and oxygen.}
\label{200_1Z1e2}
\end{figure}

In some cases, the wind temperature may increase to values of the order of $10^5\,$K
(see Fig.~\ref{200_1Z1e2}) due to friction. However, the increase is typically
smaller. Frictional heating \citep[see][]{burgers} becomes important when it is
comparable to radiative cooling
Eq.~\eqref{zaroch},
i.e.,
\begin{equation}
{n}_\text{p}{n}_\text{i}\frac{4\pi{q}_\text{p}^2{q}_\text{i}^2}{kT_{\text{pi}}}
\ln\Lambda G(x_\text{pi})\zav{v_\text{i}-v_\text{p}}=
n_\text{H}n_\text{e}\Lambda(T),
\end{equation}
where $\ln\Lambda$ is the Coulomb logarithm, $\Lambda(T)$ is the cooling function 
introduced in Eq.~\eqref{zaroch}, $T_{\text{pi}}$ is the mean temperature,
and the dimensionless velocity difference
\begin{equation}
\label{xpi}
x_{\text{pi}}=\frac{{v}_\text{i}-{v}_\text{p}}
{\sqrt{\frac{2k\zav{m_\text{i}T_\text{p}+m_\text{p}T_\text{i}}}{m_\text{i}m_\text{p}}}}.
\end{equation}
The subscripts denote values of the radial velocity $v$, number density $n$,
charge $q$, atomic mass $m$, and temperature $T$ of heavy ions (i) and passive
component (p). Using the approximations $n_\text{e}\approx{n}_\text{H}$,
$G(x_\text{pi})\approx\frac{2 x_\text{pi}}{3\sqrt{\pi}}$, $T_{\text{pi}}\approx
T$, $v_\text{i}-v_\text{p}\approx a x_\text{pi}$, approximating the passive
component by hydrogen, and using the formula in \citet{rcs} for the radiative
cooling (see also Eq.~\eqref{gozartlsk}), we can approximately determine the 
temperature
at which frictional heating is balanced to be
\begin{equation}
\label{tfric} T_\text{fric}=
9600\,\text{K}\zav{\frac{Z_\text{i}}{0.01}}^{2/3}
\zav{\frac{m_\text{H}}{m_\text{i}}}^{2/3}
\zav{\frac{{q}_\text{i}}{e}}^{4/3}\zav{\frac{{x}_\text{pi}}{0.01}}^{4/3}.
\end{equation}
We note that the a nondimensional velocity difference can be roughly calculated
using Eqs.~(18) or (23) of \citet{nlteii}.
If the wind temperature is much higher than $T_\text{fric}$, then frictional heating
is negligible. On the other hand, if the wind temperature is equal or lower than
$T_\text{fric}$, then frictional heating may influence wind temperature and
increase it to the value of about $T_\text{fric}$. Since many studied giant stars
have winds that have ${x}_\text{iH}$ of the order of $0.01$, frictional heating may
influence their wind temperature for the mass fraction of a given element
$Z_\text{i}\approx0.01$. Moreover, the effect of frictional heating is
insignificant for stars of very low metallicity $Z_\text{i}\ll0.01$.

\subsection{Decoupling of wind components}

For very low density winds, the frictional force becomes inefficient enabling
the
dynamical decoupling of wind components \citep{treni,op,kkiii,ufo}. In most
cases, the
decoupling is  stimulated by frictional heating, which leads to an
effective decrease in the frictional force (due to
the
dependence of frictional
force on
the
temperature).

\begin{figure}
\centering
\resizebox{0.9\hsize}{!}{\includegraphics{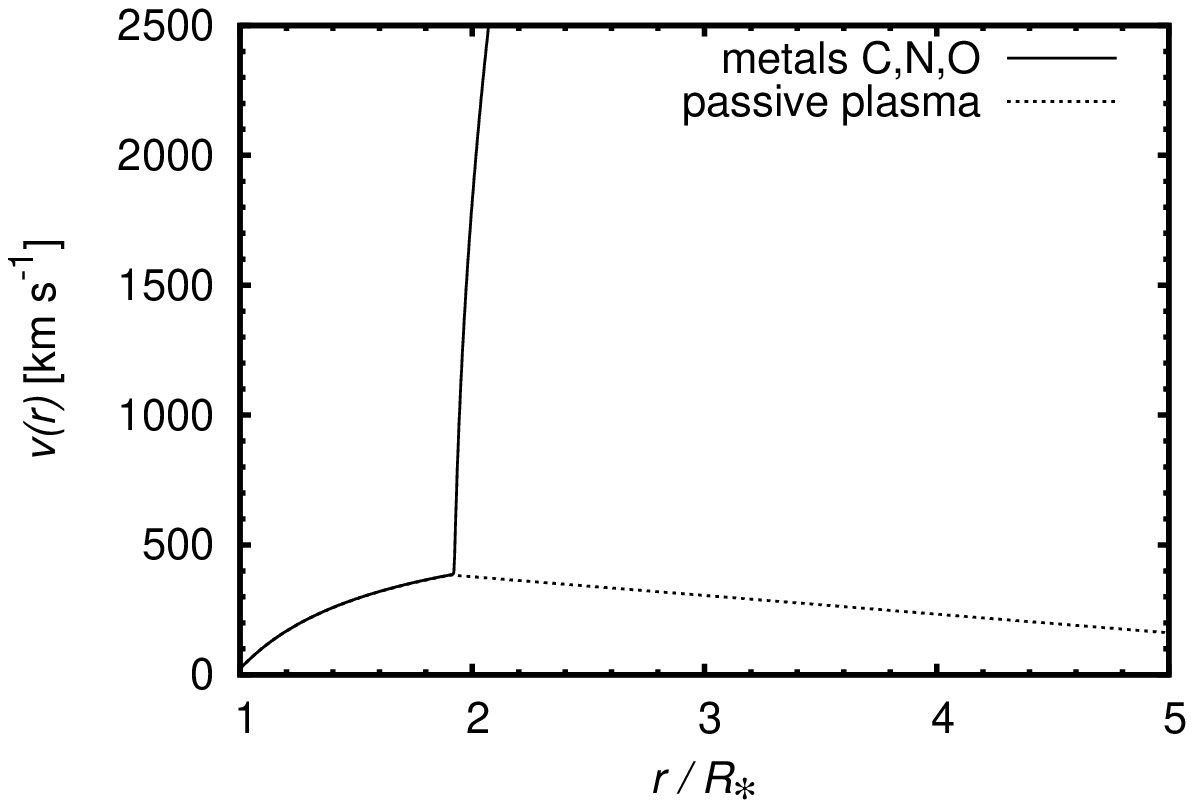}}
\resizebox{0.9\hsize}{!}{\includegraphics{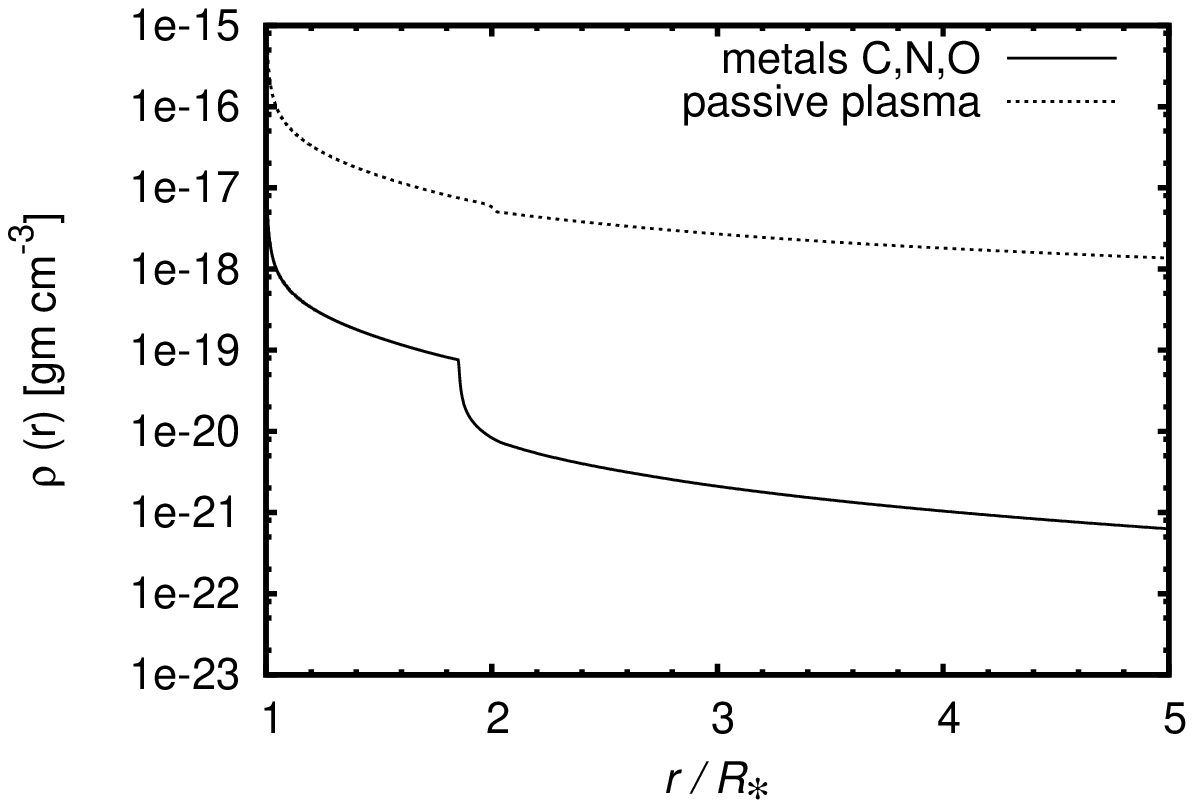}}
\caption{Calculated velocity ({\em top}) and density ({\em bottom})
profile from hydrodynamical simulations of two-component wind for the model
M100-2. Heavier ions are denoted using dashed line, hydrogen and helium using
solid line. The components decouple at $r\approx2\,\rs$.}
\label{decoupling}
\end{figure}

If decoupling occurs for velocities larger than the corresponding escape speed,
then the wind mass-loss rate remains basically unaffected by decoupling (see
Fig.~\ref{decoupling}). Here we plot results of hydrodynamical simulations of
a two-component wind for the model M100-2 (see Table~\ref{hvezpar}). The CAK constants
\citep{cak,abpar}, which characterize the effectivity of the momentum transfer from
photons to absorbing ions, are $\alpha=0.6$, $k=0.027$, and $\delta=0.1$
(corresponding to $Z=5\times10^{-3}$). At some point, the absorbing ions begin
to decouple from the passive plasma and accelerate rapidly (the so-called runaway
effect). Because this occurs at velocities larger than the escape speed, all
wind components finally leave the star. On the other hand, if decoupling were to
occur at
velocities smaller than the 
escape speed, then
the hydrogen and helium
components would be unable to leave the star \citep{obalka,kkii}. For very low
metallicities, a purely metallic wind may exist \citep{babela,uncno}.

However, the final fate of the decoupled material remains unclear, because the two-stream
instabilities may change the nature of the solution.

\begin{figure}
\centering
\resizebox{0.9\hsize}{!}{\includegraphics{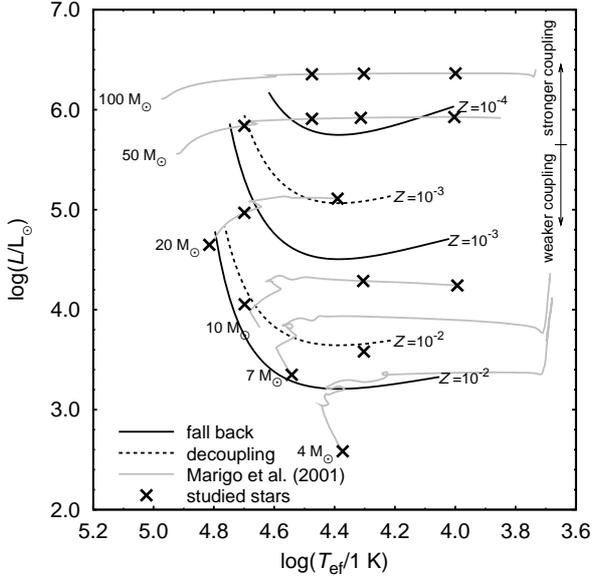}}
\caption{HR diagram for different types of stellar wind. Crosses denote studied
stars, gray lines marked by corresponding mass are evolutionary tracks
calculated by \citet{bezmari}, solid lines denote approximate location of
borders below which the fall back of hydrogen and helium occurs, and dashed
lines denote approximate location of a border below which decoupling occurs
(for a given mass fraction of heavier elements).}
\label{hrdodpad}
\end{figure}

A HR diagram showing the different types of stellar wind is given in
Fig.~\ref{hrdodpad}. For evolved stars of the highest luminosities, the
multicomponent effects are important only for very low metallicities of the
order of $Z\approx10^{-4}$. On the other hand, for less-massive stars decoupling
can occur even for a mass fraction of heavier elements comparable to the solar one
($Z\approx10^{-2}$). The shape of border lines in Fig.~\ref{hrdodpad} is given
mainly by the mass-loss rate, and partly also by the terminal velocity, wind
temperature, and charge. On average, decoupling occurs for mass fraction of CNO
elements lower than roughly
\begin{subequations}
\begin{multline}
\label{zoddel}
\log Z_\text{dec}\approx \hzav{0.48-0.60\log\zav{\frac{L}{1\,\text{L}_\odot}}}
\\*\times
\hzav{1+1.49\zav{\frac{T_\text{eff}}{10^5\,\text{K}}}-
       3.1\zav{\frac{T_\text{eff}}{10^5\,\text{K}}}^2}.
\end{multline}
For even lower metallicities, lower than
\begin{multline}
\label{zpadan}
\log Z_\text{back}\approx \hzav{0.40-0.65\log\zav{\frac{L}{1\,\text{L}_\odot}}}
\\*\times
\hzav{1+1.49\zav{\frac{T_\text{eff}}{10^5\,\text{K}}}-
       3.1\zav{\frac{T_\text{eff}}{10^5\,\text{K}}}^2},
\end{multline}
\end{subequations}
the passive component decouples from hydrogen and helium for velocities smaller than the escape speed and may
fall back onto the stellar surface.

\section{Discussion}

\subsection{Wind limits}

There exists a limiting mass-loss rate of heavier elements $\dot
M_\text{i}^\text{max}$ below which hydrogen and helium remain in the stellar
atmosphere and a purely metallic wind exists with a mass-loss rate $\dot M_\text{i}
< \dot M_\text{i}^\text{max}$. As shown by \citet{uncno}, to
achieve a
hydrostatic hydrogen solution in the atmosphere, the magnitude of the gravitational
acceleration should be larger than that of the hydrogen acceleration due to friction with
heavier elements, i.e.,
\begin{equation}
\label{vodpod}
\frac{GM}{r^2}>\frac{{n}_\text{H}{n}_\text{i}}{\rho_\text{H}}
\frac{4\pi{q}_\text{H}^2{q}_\text{i}^2}{kT_{\text{Hi}}} \ln\Lambda\,
G(x_\text{Hi}).
\end{equation}
Using the same approximations as in Eq.~\eqref{tfric} and the continuity
equation, we obtain the condition
\begin{multline}
\label{vodikat}
\dot M_\text{i} < \dot M_\text{i}^\text{max}=3\sqrt{\frac{\pi
m_\text{H}}{2}}\frac{Gk^{3/2}}{q_\text{H}^2\ln\Lambda}
\frac{m_\text{i}MT^{3/2}}{q_\text{i}^2}\\*\approx
3\times10^{-17}\,\msr\!\zav{\!\frac{m_\text{i}}{m_\text{H}}\!}\!\!
\zav{\!\frac{M}{1\,\text{M}_\odot}\!}\!\!\zav{\!\frac{T}{10^4\,\text{K}}\!}^{3/2}\!\!
\zav{\!\frac{q_\text{i}}{q_\text{H}}\!}^{-2}.
\end{multline}
Since in most considered cases in this study the derived mass-loss rate of heavier elements is
higher than $\dot M_\text{i}^\text{max}$, hydrogen (and also helium, see
Sect.~\ref{kaphelod}) are driven out of the stellar atmosphere for the
metallicities studied here. Subsequently, they may leave the star if there is no decoupling
or if decoupling occurs for velocities larger than the escape value.

For metallicities lower than those studied here, the radiative
force is, however, insufficiently strong to drive the wind containing the hydrogen and helium ions.
At these very low metallicities, purely metallic winds may occur, if
the radiative acceleration is large enough. The mass-loss rate of such a wind is lower
than that given by Eq.~\eqref{vodikat}. Since this condition does not
depend on density, we note that the value of the limiting metallic mass-loss rate $\dot
M_\text{i}^\text{max}$ is nearly constant throughout the stellar atmosphere
(with some variations caused only by the temperature and charge
variations).

\subsection{Mass-loss rate and evolutionary calculations}

From this and previous studies (\citealp*{bezvi}, \citetalias{cnovit}), we
obtain the
following picture of the stellar winds of hot first stars with an initially pure
hydrogen-helium composition.

Pure hydrogen-helium stars do not have any line-driven wind. Stars very close to
the Eddington limit with $\Gamma\gtrsim0.859$ can have very weak ($\dot
M\lesssim10^{-14}\,\msr$) pure hydrogen winds due to light scattering on free
electrons \citep{bezvi}. As a result, the influence of winds on the evolution of these
stars is negligible. In rotating stars, mass loss by means of a decretion disk (Lee et al.
1991, Ekstr\"om et al. 2008) may be of some importance. Furthermore,
massive stars may lose mass by means of either a $\eta$~Car type of explosions
\citep{vikowr,vodovar}
or a super-Eddington outflow \citep{owosha}. This outflow
may more easily exist in evolved stars as the nuclearly processed core has a lower
number of free electrons per nucleon than the hydrogen-rich envelope. 

As soon as heavier elements are synthesised in the stellar core and transported
to the stellar surface, a pure metallic wind may be produced \citep{babela} if the
radiative force on metals is large enough. These very weak stellar winds have
mass-loss rates lower than those given by Eq.~\eqref{vodikat} \citep[see
also][]{uncno}. Although this type of outflow probably does not influence stellar
evolution, it may influence the stratification of the stellar atmosphere
\citep[e.g.,][]{nedolez}.

For metallicities higher than those given by Eq.~(18) of \citetalias{cnovit},
wind that also contains hydrogen and helium may exist. For metallicities lower
than those given by Eq.~\eqref{zpadan}, the wind is very weak, wind decoupling
occurs at velocities lower than the 
escape
value, and hydrogen and helium may
fall back on the stellar surface. The final fate of such a wind is however 
unclear because realistic simulations describing their behavior are not yet
available.

For metallicities higher than those given by Eq.~\eqref{zpadan}, either winds are denser
and decoupling occurs at velocities larger than the 
escape
value or there is no
decoupling at all. The wind mass-loss rate is given by the formulae of, e.g.,
\citet{vikolamet,kudmet}, or \citetalias{cnovit}.

\subsection{High velocity particles in primordial minihaloes}

After decoupling, the CNO particles may be accelerated to velocities of the order
of $v\approx10^4\,\kms$, i.e., about $0.1c$. The energies of these particles are of
the order of $1\,\text{MeV}$, and the consequent
decoupling of the wind components
produces
the first low-energy cosmic-ray particles. The slowing-down time of
these particles in the primordial minihaloes can be considerably large. If the
interstellar material were already ionised, the slowing-down time would be
estimated to be \citep{declan}
\begin{equation}
\label{kovobrzd}
\tau_\text{s}\approx\frac{m_\text{i}m_\text{H}}{4\pi
q_\text{i}^2q_\text{H}^2n_\text{H}\ln\Lambda}v^3.
\end{equation}
Assuming that the typical hydrogen densities inside minihaloes are of the order of
$10^2\,\text{cm}^{-3}$ \citep[e.g.,][]{machacek}, the slowing-down time is of
the order of $10^5\,\text{years}$. Consequently, rapidly moving CNO particles may
easily penetrate the halo and influence its chemical composition.

The aforementioned analysis is relevant if the influence of fast metallic wind on the interstellar
medium can be neglected. In the opposite case, if acceleration of interstellar
medium is non-negligible, the medium is swept up by a metallic wind, and its density
increases.
As a result, the slowing down time \eqref{kovobrzd}
shortens, the coupling between metals and interstellar medium strengthens, and the
process repeats itself until a structure typical of wind-driven bubbles develops
\citep[e.g.,][]{kilaci}. To assess the likelihood of this scenario, we
calculate the ionized interstellar-hydrogen acceleration caused by its friction with
the metallic wind to be \citep{burgers}
\begin{equation}
g_\text{fric}=\frac{4\pi q_\text{i}^2q_\text{H}^2n_\text{i}}
{m_\text{H}k T_\text{Hi}} \ln\Lambda G(x_\text{Hi})\approx
\frac{q_\text{i}^2q_\text{H}^2\dot M_\text{i}\ln\Lambda}
{m_\text{i}m_\text{H}^2r^2v^3},
\end{equation}
where we used the approximate form of $G(x_\text{Hi})\approx1/(2x_\text{Hi}^2)$
valid for $x_\text{Hi}\gg1$ and the continuity equation of metals. Writing the
metallic mass-loss rate as $\dot M_\text{i}=\zeta\dot M_\text{i}^\text{max}$
(see Eq.~\eqref{vodikat}), the frictional acceleration is
\begin{equation}
g_\text{fric}=\frac{3\sqrt{\pi}}{4}\zeta\frac{a^3}{v^3}g(r),
\end{equation}
where we have neglected the difference in charges between atmosphere and interstellar
medium, and $g(r)$ is the magnitude of the gravity acceleration at a given
point. Although $\zeta$ may be of the order of $10^2$, the ratio of the thermal
speed in the atmosphere to the metallic wind velocity 
is $a/v\approx10^{-3}$.
Hence, the frictional acceleration of the interstellar medium is 
seven orders of magnitude lower than the gravitational acceleration and does not
significantly influence  the dynamics of the interstellar medium.

For homogeneous winds (i.e., those also including hydrogen),
both the
mass-loss rate 
%
and the ratio $a/v$ 
are larger.
In this case, the frictional force influences
the dynamics of the interstellar medium, the slowing-down time of Eq.~\eqref{kovobrzd} is
short, and the problem becomes a hydrodynamical one.

\subsection{He-free winds of cool stars?}
\label{kaphelod}

Being heavier and less fragile than neutral hydrogen, neutral helium may remain in
the stellar atmosphere, while hydrogen flows with heavier elements in the stellar
wind. Taking into account that collisions with hydrogen are more
important for helium acceleration than
collisions with heavier elements, helium remains in the stellar atmosphere if
\citep[c.f.~Eq.~\eqref{vodpod}]{my3kpoprad}
\begin{equation}
\label{helpod}
\frac{GM}{r^2}>\frac{{n}_\alpha{n}_\text{H}}{\rho_\alpha}
\frac{4\pi{q}_\alpha^2{q}_\text{H}^2}
{kT_{\text{H}\alpha}} \ln\Lambda G(x_{\text{H}\alpha}),
\end{equation}
where the subscript $\alpha$ denotes quantities corresponding to helium. 
Using the approximate formula for $G(x_{\text{H}\alpha})$ and the hydrogen
continuity equation, we 
obtain a condition for the hydrogen mass-loss rate
\begin{multline}
\label{heliat}
\dot M_\text{H} < 3\sqrt{\frac{\pi m_\text{H}}{2}}\frac{m_\alpha
Gk^{3/2}}{q_\text{H}^2\ln\Lambda}
\frac{MT^{3/2}}{q_\alpha^2}\\*\approx
10^{-16}\,\msr
\zav{\frac{M}{1\,\text{M}_\odot}}\zav{\frac{T}{10^4\,\text{K}}}^{3/2}
\zav{\frac{q_\alpha}{q_\text{H}}}^{-2}.
\end{multline}
From this, it seems that if the helium charge $q_\alpha$ is very low, i.e., if most
of the helium atoms are neutral, then helium may remain in the stellar
atmosphere,
while hydrogen and heavier elements produce a wind. The other possibility
of helium-free wind, i.e., a very low hydrogen mass-loss rate $\dot
M_\text{H}\lesssim10^{-16}\,\msr$, is not possible here because all studied stars
have mass-loss rates higher than this value.

However, if helium atoms are neutral, collisions between neutral helium atoms
and ionised hydrogen can accelerate helium into the stellar wind. Neutral helium
remains in the stellar atmosphere if the magnitude of the frictional acceleration is
smaller than the magnitude of the gravity force, i.e.,
\begin{equation}
\label{neutral}
\mu_{\text{H}\alpha}n_\text{H}n_\alpha\langle\sigma v\rangle_{\text{H}\alpha}
v_\text{H} < \frac{\rho_\alpha GM}{r^2},
\end{equation}
where $\mu_{\text{H}\alpha}=m_\text{H}m_\alpha/(m_\text{H}+m_\alpha)$ is the
reduced mass, and $\langle\sigma v\rangle_{\text{H}\alpha}$ is the momentum
transfer rate coefficient for collisions between neutral helium and protons
\citep{pigali,voda}. Adopting $\langle\sigma
v\rangle_{\text{H}\alpha}\approx10^{-9}\,\text{cm}^3\,\text{s}^{-1}$  and using
the continuity equation, the condition given by Eq.~\eqref{neutral} can be rewritten as
\begin{multline}
\dot M_\text{H} <4\pi G M \frac{m_\text{H}+m_\alpha}
{\langle\sigma v\rangle_{\text{H}\alpha}}\\*\approx
2\times10^{-13}\,\msr
\zav{\frac{M}{1\,\text{M}_\odot}}.
\end{multline}
For cooler massive stars, collisions between neutral helium and
protons alone are able to accelerate helium if the wind is strong enough.

In models of cooler stars that we have studied, there is indeed a large fraction of neutral
helium, but the charge of ionized helium is always high enough to accelerate helium
into the wind (see Eq.~\eqref{heliat}). For stars cooler than those studied here
(with $T_\text{eff}<10^{4}\,\text{K}$), collisions of neutral helium with protons
dominate the acceleration of helium in the stellar wind.

\subsection{Charging of first stars and first magnetic fields}

The problem of the escape of electrons from the stellar atmosphere was
considered by \citet[see
also \citealt{joel}]{mil}. He demonstrated that electrons, due to the radiative
forces
acting on them, can escape from the star. However, he concluded that the resulting
positive charge of the star would soon prevent any additional loss. The
problem of charging the first stars by means of their winds remains an interesting
one because of its possible implications for generation of the first magnetic fields.

Our models of the multicomponent stellar wind do not allow
us
to check the possibility
of a star increasing its charge, since the zero current condition was used to
determine the 
electron velocity at the wind base. The possible inclusion of an electron regularity
condition corrupts (although only {\em very} slightly) the zero current condition at
the upper boundary and 
can lead to the
charging of the star.

However, the physical significance of both the electron critical point and the regularity
conditions is questionable. From the hydrodynamic point of view, these conditions
can only be used if they correspond to the point where the propagation speed of
some type of disturbance is met. In general, the
characteristic equations for the multicomponent flow may not necessarily
describe the dissemination of the waves
\citep[e.g.,][]{kkiii}. Thus, hydrodynamical simulations (or at least a linear
analysis of hydrodynamical equations) should be used to prove the significance of the
electron critical point condition. Only an analysis of this type will be able to
answer the question of
whether stellar charging via stellar winds is conceivable. However, if the
possibility of stellar charging via stellar winds were proven, it could lead to
the creation of first magnetic fields in the Universe.

\subsection{The effect of wind inhomogeneities}

We have modelled hot star winds by neglecting small-scale inhomogeneities. These
inhomogeneities are expected on both theoretical \citep{ocr,felpulpal} and
observational grounds \citep{bourak,martclump,pulchuch}. However, we expect that
even in structured winds 
the inefficient transfer of momentum between the wind
components may strongly affect the wind structure. We provide at least an
estimate of the
wind parameters for which these problems may occur and we outline the inclusion
of these effects in the evolutionary calculations.

\section{Conclusions}

We have studied the effect of the 
Gayley-Owocki (Doppler)
heating and multicomponent flow
structure in CNO driven winds of hot stars. The parameters of these stars were
selected to represent massive initially pure hydrogen-helium (Pop~III) stars.

For the first time, we have included the GO heating term directly using atomic
linelist and NLTE calculations. We have shown that GO heating is important
especially for
winds of
CNO enriched first stars with high metallicities
($Z\approx0.01$).
In these winds, the
GO
heating can compete with radiative cooling 
because both
the number of strong lines
%
and the
wind 
terminal velocity 
are
large. On the other
hand, for stars with low metallicities
($Z\lesssim0.001$)
there 
are an insufficient number of
strong lines and the
adiabatic cooling dominates.

The effects of multicomponent flows are important especially at 
low metallicities
($Z\lesssim0.001$) in the case of evolved stars, and for relatively high
metallicities ($Z\approx0.01$) for main-sequence stars.
The frictional heating itself does not influence the wind mass-loss
rate. On the other hand, decoupling probably leads to a zero mass-loss rate of
hydrogen and helium if it occurs at velocities lower than the escape one. We
have developed an approximate formula that estimates the minimum metallicity above which
hydrogen and helium leave the star.

The decoupling of radiatively accelerated metals from hydrogen and helium 
leads
to generation of particles with typical energies of the order of 1\,MeV, i.e.,
the
first stars may be the first sources of low-energy cosmic rays.
%
We have shown that these particles
easily 
penetrate the
interstellar medium of a given minihalo. We have discussed the possibility of
charging
of first stars via their multicomponent winds.

Wind models presented here can also be used to describe the winds of possible
subsequent generations of CNO rich stars and low-luminosity stars of solar
chemical composition.

\begin{acknowledgements}
This research made use of NASA's ADS, and the NIST database {\sf
http://physics.nist.gov/asd3}. This work was supported by grant GA \v{C}R
205/07/0031. The Astronomical Institute Ond\v{r}ejov is supported by the project
AV0\,Z10030501.
\end{acknowledgements}

\newcommand{\actob}{Active OB-Stars:
	Laboratories for Stellar \& Circumstellar Physics, 
 	eds. S. \v{S}tefl, S. P. Owocki, \& A.~T.
        Okazaki (San Francisco: ASP Conf. Ser)}

\newcommand{\pulrot}{Pulsation, rotation, and mass loss in early-type stars,
IAUS 162, eds. L. A. Balona, H. F. Henrichs, \& J. M. Contel. (Dordrecht: Kluwer
Academic Publishers)}


\end{document}